\documentclass[epj]{svjour}
\usepackage{amsmath,amsfonts,amssymb,cite,graphics}
\begin{document}

%\preprint{UB-ECM-PF-06/21}

\title{Light meson spectrum and classical symmetries of QCD}

\author{S.S. Afonin%\inst{1}
}
\institute{Departament d'Estructura i Constituents de la
Mat\`eria and CER for Astrophysics, Particle Physics and Cosmology,
Universitat de Barcelona, 647 Diagonal, 08028, Spain. E-mail: afonin@ecm.ub.es}

\abstract{ Modern spectroscopic data on light non-strange meson
spectrum is analyzed. It is argued that the observed regularities
of experimental spectrum for highly excited states favour a
partial restoration of all approximate classical symmetries of QCD
Lagrangian (conformal, chiral and axial) broken by the quantum
corrections. The rate of restoration of classical symmetries is
estimated. The dependence of the resonance widths from the
corresponding masses is systematically checked. On average, it
turns out to be universal for the high excitations
as predicted by the effective string description.
% PACS, the Physics and Astronomy
                             % Classification Scheme.
%\keywords{meson spectrum, QCD, chiral symmetry}%Use showkeys class option if keyword
}                              %display desired

\maketitle

\section{Introduction}

The study of hadron resonances is of a great importance for a
deeper understanding of the strong interactions. As the stable
hadronic matter consists of {\it up} and {\it down} quarks, the
resonances built up of these quarks are of a special interest. It
is well known that masses of {\it up} and {\it down} quarks are
very light (of the order of 5~MeV) in comparison with typical
hadron masses (of the order 1000~MeV). Thus, with a good accuracy
one can neglect them. In this massless limit the strong
interactions are chirally invariant in the two-flavor sector. The
chiral $SU(2)_L\times SU(2)_R$ invariance is not a symmetry of the
physical vacuum. This results in the spontaneous Chiral Symmetry
Breaking (CSB) to the vector isospin subgroup $SU(2)_V$ and the
appearance of massless Goldstone bosons, the $\pi$-mesons. For
this reason the chiral symmetry is not seen in the hadronic
spectrum. The vector $\rho(770)$ and axial $a_1(1230)$ mesons
represent a typical textbook example. Another example is the axial
$U(1)_A$ symmetry broken by the chiral anomaly. This phenomenon is
known to enhance significantly the mass of $\eta'$ meson. However,
all such examples refer to ground states only, whereas the higher
radial and orbital excitations are usually avoided in QCD
textbooks.

The classical QCD action in the chiral limit has also a symmetry
with respect to the scale transformations stemming from the
absence of dimensionful constants. The scale invariance is a part
of a larger conformal group. At the quantum level this symmetry is
broken by the scale anomaly. At high energies the conformal
symmetry of QCD is, however, restored due to the asymptotic
freedom. In particular, the scaling laws of the parton model can
be derived directly from the conformal symmetry of the classical
QCD Lagrangian. In recent years it was understood that this fact
provides one with powerful tools in practical calculations: The
structure of perturbative predictions for light-cone dominated
processes reveals the underlying conformal symmetry of the QCD
Lagrangian, see~\cite{braun} for a review. This property turned
out to be crucial in connection with the conjecture about AdS/CFT
correspondence~\cite{maldac} in application to QCD, which
attracted a lot of interest recently.

The conformal symmetry is incompatible with the existence of resonances at certain energies
because the spectrum has to be scale invariant in this case. However, between the low energy
region, where the scale symmetry is badly violated, and the scale invariant high energy continuum
there is the intermediate energy region, where the resonances still exist but the conformal
invariance should be partially restored. This raises an interesting problem: How does the
partial restoration of conformal symmetry influence on the meson spectrum?

It has been suspected for long ago that QCD is dual to some string theory. The term "duality" is
commonly used when some phenomena can be described by two theories and the strong coupling regime
of one of them corresponds to the weak coupling regime of the other one. There are examples of
such duality in two-dimensional field theories.
For four dimensions duality is usually only a hypothesis, which gives, however, a powerful
tool for deriving various predictions. The large-$N_c$ limit of QCD~\cite{hoof} provides,
in a sense, a particular realization of certain duality: The theory of strongly interacting
quarks and gluons is rewritten as a theory of weakly interacting mesons and glueballs, with
baryons being the solitons in this dual theory. The two-point correlators can be then rewritten
as a sum over meson contributions,
\begin{equation}
\label{f1}
< J(p)J(-p)>=\sum_n\frac{f_n^2}{p^2-m_n^2}\sim N_c\log p^2,
\end{equation}
where $f_n=\langle0|J|n\rangle\sim\mathcal{O}(\sqrt{N_c})$ are
meson couplings. The logarithm in the r.h.s. of Eq.~\eqref{f1}
comes from the leading order of perturbation theory  (the
so-called parton model logarithm) and it is related with the
conformal invariance of classical QCD. Obviously, to reproduce
this logarithm one needs an infinite number of states, provided
the existence of confinement in the large-$N_c$. Hence, an
infinite number of narrow ($\Gamma=\mathcal{O}(1/N_c)$) meson
states is dual, at least, to the leading order of perturbation
theory, which is governed by the underlying conformal symmetry of
QCD. On the other hand, there are many arguments that QCD in the
large-$N_c$ limit is dual to some (still unknown) string theory. A
reason for such a belief is, for instance, the fact that the
planar expansion in powers of $1/N_c$ is much reminiscent of
perturbative expansion in the string theory, both expansions have
a topological nature.

Thus, QCD is believed to have some string dual and the large-$N_c$
limit strongly supports this belief. Any self-consistent string
theory possesses the conformal invariance. The approximate
conformal symmetry of QCD at the tree level gives a hope to find
this string dual, at least in some kinematic regime. But what is
then a signature of approaching to this regime for the light meson
spectrum? Probably the reply is that the spectrum should get
reminiscent of that of given by the string dual. The conformal
symmetry is crucial in making this correspondence. This gives an
idea for a mechanism of how the partial restoration of scale
invariance influences on the spectrum. The quasiclassical string
approaches typically give the following law for the light meson
spectrum: $m^2(n,J)\sim n+J$, where $n$ is the principal (radial)
quantum number and $J$ is the spin. This spectrum is provided by
the Veneziano type of strings. Hence, the QCD string dual,
probably, represents a some modification of {\it \`a la} Veneziano
string with a similar spectrum. A special feature of this spectrum
is that it predicts the clustering of states with different $n$
and $J$ near certain equidistant values of masses squared defined
by the sum $n+J$. Thus, if one experimentally observes a tendency
to such a clustering, this phenomenon could be interpreted as a
manifestation of the partial restoration of conformal invariance
of the underlying fundamental theory.

In principle, the corresponding physics can be figured out without
exploiting the string ideas. If some 'rest' of conformal symmetry
is indeed realized in the meson spectrum, then the physical states
must fill out the corresponding group representations with
degenerate masses inside one multiplet. Experimentally these
multiplets should be then observed as clusters of states near some
values of energy. Unfortunately, it seems that nothing is known on
this subject.
%For this one should explain what is the physical sense of the partially realized scale symmetry
%in some energy %region. The physical sense is that the given symmetry is realized everywhere in
%this region except some points.
%If the resonances exist in this energy region then their masses have to cluster near these
%points.

%Consequently, if this happens, the observed pattern of regularities in the spectrum has to
%be caused by a %partial restoration of conformal symmetry of classical QCD Lagrangian. It may
%even happen that the string dual %does not exist in reality, but the conclusion remains intact.

%These phenomena, in particular, strongly hamper the formulation of QCD
%as a consistent string theory.
Motivated by these discussions, in
the present paper we address to the problem of relations between the experimental
spectrum of light non-strange mesons and approximate classical
symmetries of QCD broken at the quantum level.

The paper is organized as follows. The details of phenomenological analysis are given in
Section~2. Sections~3 and~4 are devoted to the interpretation of observed regularities for
the masses and decay widths correspondingly.
We conclude in Section~5.

\section{Experimental spectrum}

The radial and orbital excitations were only poorly known in the time
of fast development of QCD in 70's. Since that time the experimental
data has been accumulating and now Particle Data Group (PDG)~\cite{pdg}
lists a certain number of well established higher excitations in the
light non-strange meson sector (we will denote these states $\bar nn$)
up to the energy of 1.9~GeV. At higher energies PDG
enumerates only a few confirmed mesons and many unconfirmed states.
At present it is difficult to draw any direct conclusions about
general properties of meson excitations based on the well confirmed
states of PDG only. To reveal these properties we propose an indirect
way: together with well confirmed states one can analyse many
unconfirmed (more precisely, not well confirmed) states.
As it usually happens in a large statistical ensemble, one can hope
that possible errors in different channels smooth each other providing
finally a stable general picture, which can be described by some mean
characteristics.

Since PDG cites so many unconfirmed states it is easy
to go astray in searching for regularities. To avoid this one
inevitably should stick to some reasonable principles.
Let us explain how we choose the unconfirmed resonances for
the analysis. First, for reliability we will take only those
states which were observed at least in two different reactions.
Thus, we will deal with the 'not well confirmed' mesons rather than
with the 'unconfirmed' ones. Second, at energy above 1.9~GeV we will
use the data of the Crystal Barrel Collaboration on proton-antiproton
($\bar pp$) annihilation in flight. The latest review of this
data is contained in ref.~\cite{bugg}. The
reasons for this choice are as follows:\\
\begin{enumerate}
\item
It is the only experiment which performed a systematic study
of the mass range 1.9-2.4 GeV. The coverage of this mass range from
other experiments is very limited.
\item
As a rule the states were independently observed in different
channels, {\it i.e.} they are quite reliable. The reason why most of
them are listed by PDG in a section 'Other Light Unflavoured
Mesons' is that PDG requires confirmation from a separate
experiment. The appearance of other states in this
section usually has a rather sporadic character.
\item
As it was realized long ago~\cite{barg}, meson resonances are
strongly coupled to the $\bar NN$ reactions because mesons have the quantum
numbers of the $\bar NN$ system. The dominant role of this system in the
dynamics of meson states makes the data extracted from the $\bar NN$ reactions
quite reliable.
\item
A possible admixture of strange quarks is a serious problem
for any classification of light states. A feature of $\bar pp$ annihilation
is that the production of $\bar ss$ component is strongly suppressed.
Consequently, it is quite reliable that the discovered states are, except
some rare cases, genuine $\bar nn$ mesons.
\item
The obtained spectrum (first systematized in ref.~\cite{ani}) turned out
to be in a full agreement with the old theoretical expectations from
the hadron string models and the low-energy amplitudes~\cite{LS}. Namely,
\begin{enumerate}
\item
linearity of Regge trajectories;
\item
equidistance of daughter Regge
trajectories (linearity of radial Regge trajectories), as a consequence
approximate
universality of slopes of trajectories;
\item
the intercept of the pion Regge
trajectory is approximately equal to 0 and the intercept of the $\rho$-meson
one is 0.5;
\item
the slope of the radial Regge trajectories is about $2m_{\rho}^2$
which is also consistent with the lattice calculations (see, {\it e.g.},~\cite{kac}).
\end{enumerate}
\item
To reveal the general properties of spectrum it is preferable to use the
data of an individual systematic experiment. Only after that the overall picture
should be compared with the one given by another systematic experiment.
This type of comparison can lead to some global shifts but does not spoil the picture
qualitatively.
If one first performs the data averaging (as it is done by PDG), the errors
accumulate rapidly and the final picture can be completely obscured.
The case of light non-strange baryons is a good example: If one separately
uses the data of individual systematic experiments (say, going under the names
"Cutkosky" or "Hoehler" in PDG)
the multispin-parity clusters of states are unambiguously seen, but if one
takes the averaged data of PDG the clustering gets rather controversial.
\end{enumerate}

After these general arguments let us pass on to the analysis. As
said above, we do not consider the states with a large admixture
of strange quarks (usually it is clear from analysis of
corresponding decay channels) and we omit all states which were
observed in one channel only (although many of them fill well the
missing states on meson trajectories). In review~\cite{bugg} the
latter states are: $\omega(2205)$, $a_1(1930)$, $a_1(2270)$,
$a_2(1950)$, $a_2(2175)$, $\omega_4(2250)$, $b_5(2500)$\, and
$f_6(2485)$. For the same reasons we omit $h_1(1595)$ and
$b_1(1620)$ (see~\cite{bugg} for references). We do not use
$\pi_2(1880)$ and $\eta_2(1870)$ which were cited in~\cite{bugg}
and were shown to be inconsistent with $\bar nn$ state. Similarly
we omit $f_0(2100)$ from~\cite{bugg}, which is either a glueball
or $\bar ss$ state strongly mixed with $\bar nn$. The very narrow
$\rho(1900)$ cited by PDG~\cite{pdg} (in the list of unconfirmed
states) is also not considered since there are many doubts that it
is a real resonance. The only well confirmed states of PDG which
are exotic for the quark model are $\pi_1$ mesons, namely
$\pi_1(1400)$ and $\pi_1(1600)$. The state $\pi_1(2015)$ was seen
in two reactions. We decided to include them into analysis because
at least the first two of them are generally recognized observable
resonances. We also include $f_0(980)$ and $a_0(980)$ although the
nature of these states is still controversial, presumably they
have a large admixture of strange quark (see, {\it e.g.}, note on the
scalar mesons in ref.~\cite{pdg}). The reason will be explained below.
The state $\eta(547)$ has a large admixture of strange component.
Nevertheless, this admixture does not seem to be dominant in the
corresponding radial excitations. For this reason $\eta$-meson is
also considered.

Let us explain how we display the data. First, in the relativistic theories one
deals with (masses)$^2$ which appear in the multiplets, Regge and string theory
{\it etc.}, and only these quantities are of theoretical interest.
Second, it is better to normalize all masses to some typical hadron mass.
In our opinion, the best candidate for the normalization is the mass of $\rho$ meson.

The final picture of meson spectrum resulting from our analysis is displayed in Fig.~1.
The corresponding experimental data is given in Table~1.

A well-pronounced feature of the spectrum is that the observed states cluster
about some values of energy~\cite{bugg,als}. A similar phenomenon exists in the
light non-strange
baryons~\cite{hol}. The clustering occurs at approximately 1.33, 1.70, 2.00 and
2.27~GeV. Some channels have additional states denoted by open circles or
strips in Fig.~1. They appear because the states in these channels can be created
by different orbital momenta which results in doubling of the corresponding radial
Regge trajectories~\cite{ani}. For $\rho$ and $f_2$ mesons there is polarization
data which separates S-wave from D-wave and P-wave from F-wave states
correspondingly~\cite{bugg}. For other channels such separation is tentative and
new experiments are called for.

For the clusters we display in Table~1 the mean mass $\bar M$ and the mean full decay
width $\bar \Gamma$, which are defined as follows,
\begin{equation}
\bar M \equiv\sqrt{\frac1k\sum_k m^2_k},\qquad
\bar \Gamma \equiv\frac1k\sum_k \Gamma_k,
\end{equation}
where the index $k$ enumerates the states in a cluster. The rules for the averaging are natural:
The observable quantities are $m^2_k$ (as discussed above) and $\Gamma_k$. The data for $\bar M$
and
$\bar \Gamma$ is presented in the form ('m.' denotes 'mean')
$$
\bar M,\bar \Gamma= \text{m. value}\pm\text{m. square deviation}\pm\text{m. exper. error}
$$

It must be emphasized that the positions of clusters are very stable due to
many states involved. For instance, above 1.9~GeV one can consider only those
states from~\cite{bugg} which have the maximal star rating (rating $4^*$ according
to the classification in~\cite{bugg}). These resonances require observation of
3 or more strong, unmistakable peaks and a very good mass determination. Their
reliability is practically equivalent to that of states in PDG. There exist 6 such states
in the third cluster and 8 in the fourth one. One can check that if we
consider only these states in the clusters, the positions
of the clusters will not change ({\it i.e.} the change will be less than 0.01 GeV within
our accuracy).

The clusters describe the behavior of the spectrum as a whole (the relevant
discussions for the light non-strange baryons can be found in ref.~\cite{ki}). With
a good accuracy they are equidistant, hence, one can parametrize them by a
linear function. For the data in Fig.~1 the result of the fit is
\begin{equation}
\label{1}
M^2(n)=an+b, \,\,\, n=1,2,3,4; \quad a\approx1.13,\, b\approx0.63,
\end{equation}
where $M^2(n)$ is the position of the $n$-th cluster in GeV$^2$.
The slope $a$ in cluster spectrum~\eqref{1} is nothing but the mean slope of
radial Regge trajectories. Its numerical value is within the interval
found in~\cite{bugg}: $a=1.14\pm0.013$. The parameter $b$ is the mean
intercept of the radial Regge trajectories. In~\cite{bugg,ani} this quantity
was not estimated, but for us it is of importance as will be seen below.

Finally we would like to estimate to what extend the cluster spectrum in Eq.~\eqref{1}
is vitiated if one excludes the Crystal Barrel data (the last two clusters). Then we have
only two clear-cut clusters. The parametrization of two points by the linear function can
look doubtful, so we will consider the ground $\rho$ and $\omega$ mesons as two non-strange
constituents of the lowest cluster near $0.78$~GeV. We note in passing that fit~\eqref{1}
predicts
the lowest cluster for $n=0$ near $0.79$~GeV, so our assumption is well justified. We have then
\begin{equation}
\label{1.2}
M^2_{\text{\tiny PDG}}(n)=an+b,\,\,\, n=0,1,2; \quad a\approx1.14,\, b\approx0.61.
\end{equation}
Both cluster spectra \eqref{1} and \eqref{1.2} turn out to be very close. 
Thus, PDG contains enough data to arrive at
our conclusions. The data of Crystal Barrel provides a dramatic confirmation for the observed
regularities.

Concluding this section we would like to make the following remark.
The hypothesis that mesons should appear as towers of states
(which we call clusters, Fig.~1 is selfexplanatory in the analogy
with towers) was proposed before QCD~\cite{barg} for explaining
the absence of backward peaks in $\pi^+\pi^-$, $\pi^+K^-$, $K^+K^-$,
and $\bar NN$ elastic scattering in the framework of Regge theory.
There was a hope that further $\bar NN$ studies
(see reason~3 above)
would provide crucial tests for the existence of these towers.
The Crystal Barrel experiment on $\bar pp$ annihilation can be
considered as such a test.
%In view of the modern data it would be
%interesting to reconsider the old models based on Regge phenomenology
%to check their range of validity.

\section{Interpretation of data}

It is well known that properties of any quantum system approach to
its classical ones while the quantum numbers defining the
stationary states of this system are large enough (see,
{\it e.g.},~\cite{landau}). In our case these quantum numbers are the
spin $J$ and the radial excitation number $n$. The valent quarks in
such hadrons on average have high energies and, hence, practically
do not 'feel' the non-per\-tur\-ba\-tive structure of QCD vacuum which
is the underlying reason of CSB (see the relevant discussions
in~\cite{sh} and references therein). The quasiclassical description
implies, in particular, the universal linear string-like behavior of
meson mass spectrum which can be compactly written as
\begin{equation}
\label{2}
m^2(I,G,P,C,L,J,n)\simeq a(n+J)+b.
\end{equation}
The fact that relation~\eqref{1} is an experimental result and
that the number of states is indeed growing in clusters seems to
confirm the validity of the quasiclassical treatment.
Needless to say that the manifest cluster structure of high meson
excitations includes the full linearly realized approximate
$U(2)\times U(2)$ chiral flavor symmetry of QCD as a particular
case. The restoration of this symmetry at high energies leads
to degeneracy inside the chiral multiplets.
Different aspects of the relation between the chiral symmetry restoration
in highly excited hadrons and the parity doubling were widely discussed in
the literature~\cite{als,sh,many,G3,we2,we,we3}.
It happens, however, that with the same accuracy
the observed mass degeneracy is much
higher than predicted by the restoration of chiral and axial
symmetries of
QCD Lagrangian. Even models of the generalized chiral symmetry
like in ref.~\cite{G3} cannot explain such a high degeneracy because
in the chiral multiplets one has the states with equal spin only.
This phenomenon should be a manifestation of some additional
symmetry. If we believe that all regularities in the spectrum
must be related to the symmetries of QCD, then we have only one
possible candidate: the conformal symmetry.
According to the general principle in the quantum
theories discussed at the beginning of this section, one expects
the restoration of {\it all} broken classical symmetries of QCD Lagrangian
in highly excited states.
The conformal invariance is among the classical symmetries. Consequently,
it should be effectively restored at high energies. Indeed, we
know that QCD is nearly conformal in the ultraviolet region.
As discussed briefly in Introduction, this could be intimately related with
the existence of string dual for QCD. However,
following the arguments given in Introduction,
even without this duality it is natural
to suggest that {\it the observed degeneracy of the light
non-strange mesons is a combined effect of a partial
restoration of chiral and conformal symmetries at high energies}.

Unlike the case of chiral invariance, where the complete restoration
is possible in the spectrum ({\it i.e.} the complete parity doubling),
we cannot observe in the spectrum the complete restoration of conformal
invariance, which could mean the ideal degeneracy of states with
different spin inside a cluster, like in the string theories of Veneziano type.
In QCD the existence of hadrons at discrete energies is incompatible
with the absence of scale in the problem. What we can observe is only
approaching to that regime. Finally the resonances disappear and the
scale invariant continuum sets in. Thus, the approaching to the
perturbative continuum and the grouping of resonances into clusters
seem to be tightly related.

As seen qualitatively from Fig.~1 and numerically from Table~1 the higher resonances one
considers the more clear-cut clusters they form. Let us estimate the rate of clustering.
In doing this certain care should be exercised. This procedure makes sense only if the
deviations from the averaged values are substantially larger than the corresponding experimental
errors. As seen from Table~1 this is indeed the case for the first and second clusters, where the
deviations are by a factor of $3\div4$ larger than the averaged experimental errors. For the third
cluster the difference is by a factor of 2 only, while for the fourth one there is practically no
difference at all. Thus, only the first two clusters (the PDG data) can serve for our purpose
more or less reliably. The Crystal Barrel data will be used for a qualitative check.

The equidistant cluster spectrum with deviations can be written in the form
\begin{equation}
M(n)=\sqrt{an+b}\pm\delta(n).
\end{equation}
Now we should interpolate the deviation $\delta(n)$ by some smooth function. {\it A priori}
we have no theoretical idea how this function should look like. In the literature there exist
some arguments for the rate of chiral symmetry restoration only. Namely, in~\cite{we2,we3} the
deviations were argued to be exponential, while in~\cite{sh} a polynomial minimal rate was
derived. We will consider the both possibilities for $\delta(n)$,
\begin{equation}
\delta_{\text{e}}(n)\sim\frac{e^{-\beta_{\text{e}}n}}{\sqrt{n+1}},\qquad
\delta_{\text{p}}(n)\sim(n+1)^{-\beta_{\text{p}}n}.
\end{equation}
Here in the first ansatz we introduced the square root in order to have a purely exponential
correction for the mass squared. Taking the corresponding values for the first two clusters
in Table~1, $\delta(1)\approx89$~MeV,
$\delta(2)\approx56$~MeV, one arrives at the following estimates
\begin{equation}
\beta_{\text{e}}\approx0.26,\qquad \beta_{\text{p}}\approx1.14.
\end{equation}
The resulting predictions for the exponential and polynomial deviations are (in MeV):
$\delta_{\text{e}}(3)\approx37$, $\delta_{\text{e}}(4)\approx26$, $\delta_{\text{p}}(3)\approx40$,
$\delta_{\text{p}}(4)\approx31$, while experimentally $\delta(3)\approx40$, $\delta(4)\approx37$.
It is seen that the polynomial ansatz works slightly better. The modern level of experimental
accuracy, however, does not allow to indicate convincingly which ansatz is really preferable.

Let us speculate about the physical sense of these estimates. If the partial restoration of
conformal invariance indeed takes place then the obtained results can be considered as a rough
estimate for the rate of this restoration. On the other hand, they may be regarded as an estimate
for the minimal rate of the chiral symmetry restoration in excited hadrons. Obviously, in
particular channels this effect can occur faster. For instance, the fits in~\cite{we2} yield
$\beta_{\text{e}}\approx1$, while for the polynomial ansatz there exist the estimate for the
scalar channels~\cite{sh}, $\beta_{\text{p}}\gtrsim1.5$. In any case it should be noted that
although the estimation of this rate is still a rather controversial problem, the effect of
chiral symmetry restoration in highly excited states {\it per se} seems to be now well settled
both experimentally and theoretically.

%The use of quasiclassical arguments may look only suggestive.
In principle, one can
try to give some alternative explanations of experimental data, say,
for parity doubling. For instance, it may be that
at high energies the influence of CSB ceases to depend on the
quantum numbers of concrete channel, but the strong CSB persists at
all energies, just higher excitations 'feel' its presence equally.
Or it may be that the mass degeneracy is just an effect of vanishing
spin-orbit forces in quark interactions as it was proposed
for light baryons in ref.~\cite{klempt}. Some
independent tests are needed.

Actually, such a test can be provided by the results of some recent papers.
Due to the observed
universality of spectra expressed by existence of
clusters~\eqref{1} it seems to be sufficient to check the
situation for some channels only. The channels with $J=1$
are good candidates for our purpose because
the problem can be directly addressed in these cases within
the QCD sum rules in the planar limit. Since the spectrum is
linear with a good accuracy,
one can saturate the sum rules by the linear ansatz for the masses.
As it was noted in~\cite{we2,we} and developed in~\cite{we3},
if the chiral symmetry is not broken (which is equivalent to the
absence of the weak pion decay constant and the quark condensate)
then the linear spectrum turns out to be: $M^2(n)=a(n+1/2)$.
For $J=1$ mesons this relation
is a consequence of absence of gauge-invariant local dimension-two
gluon condensate. If the CSB is present at
any energy then the intercept should be substantially larger,
{\it e.g.}, in ref.~\cite{we} it was obtained for this case
$M^2(n)=a(n+1)$. Experimentally one has for $J=1$ clusters:
$m^2(n)\approx a(n+0.51)$ with $a\approx1.13$ GeV$^2$. Consequently,
experimental spectrum on average
favors the chirally symmetric pattern, {\it i.e.} it reveals a strong
supression of CSB effects for high excitations where the linear
behavior sets in. For experimental spectrum~\eqref{1} one obtains
$m^2(n)\approx a(n+0.55)$ with the same slope. Thus, the
universality works remarkably well providing a solid ground for
extension of the conclusion to the whole spectrum.

Finally let us consider the states $f_0(980)$ and $a_0(980)$ the
nature of which is a subject of many discussions in the literature. A large
amount of phenomenological arguments (see ref.~\cite{animat} and the
references therein) indicates that these mesons are genuine $\bar
nn$ states with a large admixture of $\bar ss$ component which
shifts their masses almost to the $\bar KK$ threshold. The
analysis of different reactions (see~\cite{animat} for references)
yields the estimation of strange component in $f_0(980)$ to be
about 60-70\%. In fact, this estimate can be easily obtained
theoretically. After the CSB the ground vector-isovector meson and
the ground scalar-isoscalar meson practically do not mix~\cite{gh}. If
the latter state is $f_0(980)$ then we should have
$m_{\rho}^2=m_{f_0}^2$. Since $\rho$ meson is a pure $\bar nn$
state, the estimation of $\bar ss$ admixture in $f_0(980)$ follows
immediately: $(m_{f_0}^2-m_{\rho}^2)/m_{\rho}^2\approx0.6$ (we
remind that in all relevant formulae one deals with (masses)$^2$).
The situation with $a_0(980)$ happens to be similar. Such an
estimation does not give any insight into a mechanism of this
admixture and it may be that for $f_0(980)$ and $a_0(980)$ this mechanism is
different. It only shows that numerically it is consistent with
the hypothesis that these mesons are genuine quark-antiquark
ground scalar states. Thus, if the strange quarks were 'switched
off' the lowest cluster at about 0.78~GeV would consist of four mesons:
$\rho(770)$, $\omega(782)$, $f_0(980)$ and $a_0(980)$
(that is why the last two particles have been included into our analysis).
As noted above, this cluster is in a good agreement with spectrum~\eqref{1}
for $n=0$.

\section{Analysis of decay widths}

The clusters of meson states have not only the stable positions at some equidistant values of
energy,
but also the stable mean decay widths. They are shown in Table~1. What do they tell us about?
In the given section we address to this question.

It is widely believed that the light mesons can be considered as an effective hadron string
with relativistic quarks at the ends. The conjecture is that a flux tube of the chromoelectric
field between a quark and an antiquark can be effectively described as a string. On the basis
of such a simple qualitative picture the following behaviour for the full decay width was
predicted~\cite{cnn}: $\Gamma(n)=Bm(n)$, with $B=\mathcal{O}(1/N_c)$ being a universal constant.
Originally this relation was derived for highly excited states, where a quasiclassical treatment
can be applied. With this result at hand, let us consider the behaviour of the mean width in the
clusters, namely introduce the number $B(N)$ defined as,
\begin{equation}
B(N)\equiv\frac{\bar\Gamma(N)}{\bar M(N)}.
\end{equation}
For $N=1,2,3,4$ the corresponding values are given in Table~1.

It is desirable to have an estimate for $B(0)$ as well. Here one must exercise certain care
because the averaging of widths for the ground states should not be the same as for the excited
ones. First of all, we do not consider
the mesons $f_0(980)$ and $a_0(980)$ because a large admixture of the strange quark in these
states is expected to change dramatically their widths. Although above we have presented an
argument why these states can be considered as the members of the lowest cluster in the analysis
of the mass spectrum, this hardly can be done for the widths. Second, the full width of the
$\omega(782)$-meson, $\Gamma=8.49\pm0.08$ MeV, is almost by 18 times less than that of
$\rho(770)$-meson, $\Gamma=150.3\pm1.6$ MeV. The flavour symmetry predicts an approximate
mass degeneracy for these states, but this is emphatically not the case for their widths.
The reason is that the decay
$\omega\rightarrow\pi\pi$ is strongly suppressed for the flavour singlet and the dominant
decay is $\omega\rightarrow\pi\pi\pi$. The latter has much less phase space. Its first radial
excitation, the
$\omega(1420)$-meson, avoids this by decaying into $\rho\pi$. This phenomenon has nothing to
do with our subject and, hence, we have to exclude the $\omega$-meson from the averaging.
Finally, we have only the $\rho$-meson, which gives
\begin{equation}
B(0)=\Gamma_{\rho}/m_{\rho}=0.194.
\end{equation}

It is interesting to note that this number was proposed in~\cite{sh1} as an educated guess in
order to estimate the constant $B$ in the real world. Surprisingly enough, it turned out to be
very close to a numerical estimate
for $B$ in the 't Hooft model (QCD in two dimensions in the large-$N_c$ limit, see~\cite{dim2})
performed in~\cite{sh2}.

Finally, our analysis yields the following estimates (see Table~1)
\begin{equation}
\label{stab}
B(0)\approx B(1)\approx0.2, \,\quad B(2)\approx B(3)\approx B(4)\approx 0.1.
\end{equation}
Thus, the educated guess made in~\cite{sh1} turns out to be correct for the next cluster ($N=1$)
as well.
However, then one observes a sudden jump down by about 2 times. The Crystal Barrel Data, which was
used to estimate $B(3)$ and $B(4)$, dramatically confirms this jump. How can we interpret this
phenomenon? Looking at the states in clusters more attentively, one can make the following
observations: (a) the $N=1$ cluster mainly consists of the ground states; (b) the mesons in
the $N=0,1$ clusters prefer to decay into two particles~\cite{pdg}; (c) the mesons in the $N>1$
clusters prefer to decay into three or four particles~\cite{pdg}. The nature of the phenomenon (c)
is enigmatic, at least for the author. This very phenomenon leads to a suppression of the
available phase space for the decays. But how can one understand this within the effective
hadron string? In the simplest case of open string one could assume, for instance, that the
string breaks in two points simultaneously, producing three final particles. This is a
$\mathcal{O}(1/N_c^2)$ effect. The question arises, why this effect might become dominant?
A more plausible assumption is that the string decay is a cascade process for the excited states.
Experimentally one cannot detect the intermediate stages of this process. The universal quantity
$B$ is somehow decreased in this case.

Thus, the stability of numbers in Eq.~\eqref{stab} supports the possibility of the effective
string description. However, the experiment seems to tell us that this description for the excited
states should be different from that of the ground states, at least with respect to the issue of
string decays.

Last but not least. If one considers the individual channels, the result~\eqref{stab} hardly
can be detected. In this respect the t'~Hooft model provides an instructive example. It has so
little degrees of freedom for exciting the bound states that each cluster consists of one state
only. As a result, the quantity $B(n)$ has seemingly random fluctuations around a constant value,
which is clearly seen when one computes the widths for several hundreds of radial
excitations~\cite{sh1,sh2}. Dealing with the first several states, this asymptotic value cannot
be guessed at all. In four dimensions we are more lucky. The multitude of states in each cluster
smooth significantly these fluctuations after the averaging. Due to this effect already a few
clusters are able to provide the asymptotic value for B.

\section{Conclusions}

Our analysis shows that the available experimental spectrum of light non-strange
mesons reveals the universal string-like behavior expressed by
Eq.~\eqref{2} and, on average, a strong suppression of chiral symmetry
breaking effects for sufficiently high resonances.
%Consequently, these two
%requirements should be satisfied in any model pretending to describe
%the whole light meson spectrum.
The observed degeneracy of the spectrum, however, cannot be explained
by effective restoration of the chiral and axial symmetries only.
A possible explanation is that a partial restoration of the conformal
invariance happens simultaneously.

Independently of interpretation, the modern experimental data seems to point out 
two remarkable facts~\cite{rem}, which hold on average for the high
excitations of light non-strange mesons. First, the spectrum globally
behaves as that of the Lovelace-Shapiro dual amplitude 
(the intercept is the half of the slope). Second, the full decay width is 
proportional to the the mass of decaying particle, just as expected
within various string models.  

New systematic experiments for the search of light non-strange hadrons
above 1.9~GeV are indispensable. Unfortunately, at present such
experiments are not very wi\-de\-sp\-read because they are not expected to
bring a new physics. The analysis carried out in the paper is, in a sense,
an attempt to overcome this prejudice. The detailed knowledge of experimental
spectrum for the high meson excitations can help significantly to answer some
fundamental questions and, hence, to extend our understanding of QCD.

\section*{Acknowledgements}

I am grateful to D.V. Bugg for useful communication concerning the
modern experimental situation, to H. $\check{\text{S}}$te\-fan\-$\check{\text{c}}$i\'c
for critical reading of the manuscript at the first stage of the work, to
J. Russo and A. Cotrone for discussion about the string decays and to M. Shifman and A.
Vainstein for various useful discussions.
The work was supported by
CYT FPA, grant 2004-04582-C02-01, CIRIT GC, grant 2001SGR-00065,
RFBR, grant 05-02-17477, and by Ministry of Education and Science
of Spain.

%\begin{center}
\begin{figure*}
\label{fig}
\vspace{-4cm}
\hspace{-2cm}
%\vspace*{5cm}
\resizebox{1.2\textwidth}{!}{%
  \includegraphics{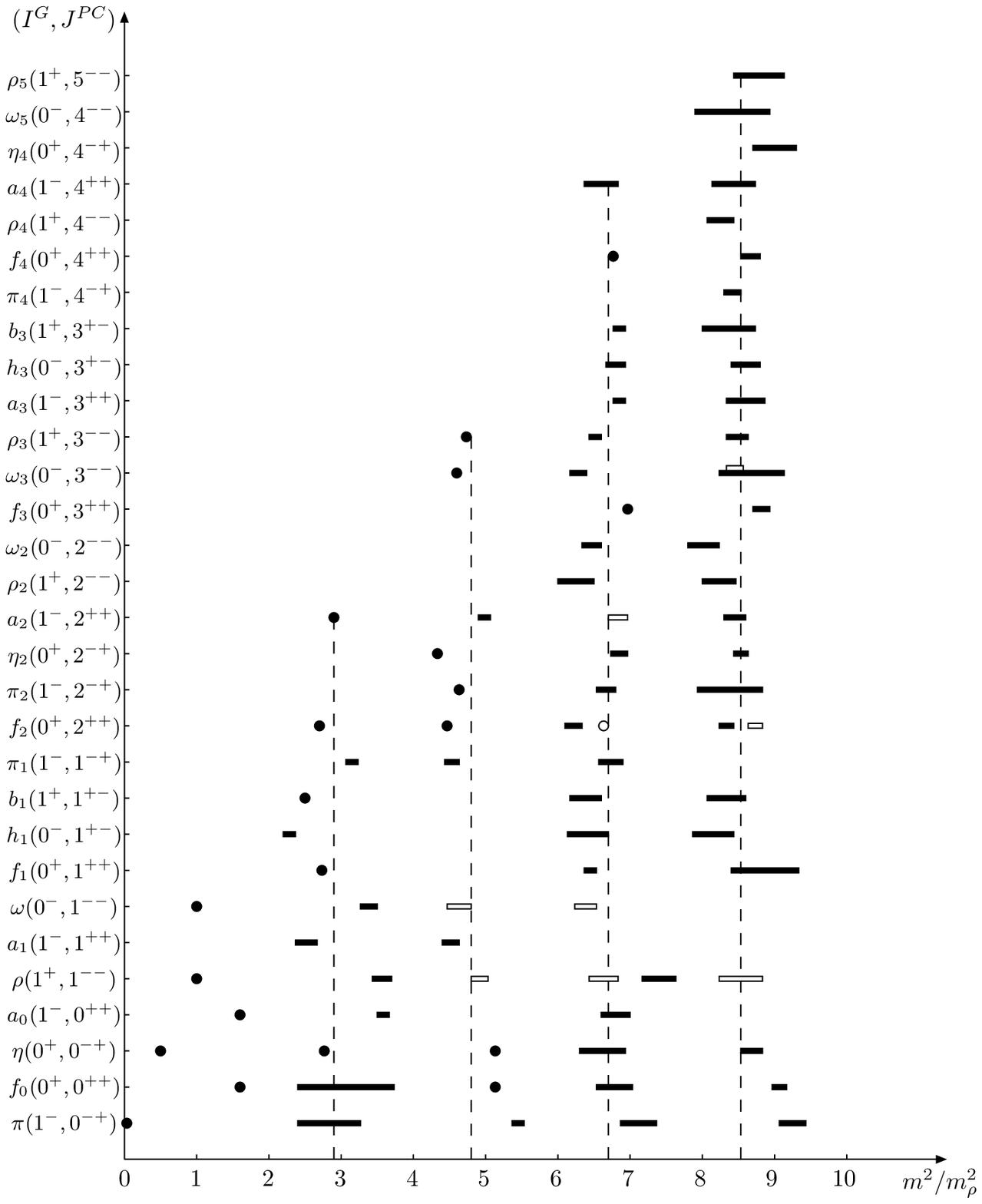}
}
\vspace{-7.5cm}
\caption{The spectrum of light non-strange mesons from refs.~\cite{pdg} and~\cite{bugg} (for the last two clusters)
in units of $m_{\rho(770)}^2$. Experimental errors are indicated. Circles stay when errors are negligible.
Open circles and strips denote the additional states (see text). The dashed lines mark the mean (mass)$^2$
in each cluster. The absolute values of masses are given in Table~1. The masses of the lightest states not
displayed in Table~1 are (in MeV):\quad
$\pi$: $140$;\quad
$f_0$: $980\pm10$;\quad
$\eta$: $547.75\pm0.12$;\quad
$ a_0$: $984.7\pm1.2$;\quad
$\rho$: $775.8\pm0.5$;\quad
$\omega$: $782.59\pm0.11$.
}
\end{figure*}
%\end{center}

%\vspace{-9.3mm}

\begin{table*}
\caption{The masses and widths (in MeV) of states in Fig.~1. Experimental errors are indicated.}
\label{tab:1}       % Give a unique label
% For LaTeX tables use
\begin{tabular}{llllllllll}
\hline\noalign{\smallskip}
 & %$M(0)$ & $\Gamma(0)$ &
 $m(1)$ & $\Gamma(1)$ & $m(2)$ & $\Gamma(2)$ & $m(3)$ & $\Gamma(3)$
& $m(4)$ & $\Gamma(4)$  \\
\noalign{\smallskip}\hline\noalign{\smallskip}
$\pi$& %$140$&&
$1300\pm100$&$200-600$& $1812\pm14$&$207\pm13$& $2070\pm35$&$310^{+100}_{-50}$& $2360\pm25$&$310^{+100}_{-50}$\\
$f_0$& %$980\pm10$&$40-100$&
$1200-1500$&$200-500$& $1770\pm12$&$220\pm40$& $2020\pm38$&$405\pm40$& $2337\pm14$&$217\pm33$\\
$\eta$& %$547.75\pm0.12$&&
$1294\pm4$&$55\pm5$& $1760\pm11$&$60\pm16$& $2010^{+35}_{-60}$&$270\pm60$& $2285\pm20$&$325\pm30$\\
$a_0$& %$984.7\pm1.2$&$50-100$&
$1474\pm19$&$265\pm13$&&& $2025\pm30$&$300\pm25$&&\\
$\rho$& %$775.8\pm0.5$&$150.3\pm1.6$&
$1465\pm25$&$400\pm60$& $1720\pm20$&$250\pm100$& \!\!\!\!\!\!
\begin{tabular}{l}
$2000\pm30$\\
$2110\pm35$\\
\end{tabular}&\!\!\!\!\!\!
\begin{tabular}{l}
$260\pm45$\\
$230\pm50$\\
\end{tabular}&
$2265\pm40$&$325\pm80$\\
$a_1$& $1230\pm40$&$250-600$& $1647\pm22$&$254\pm27$&&&&\\
$\omega$& %$782.59\pm0.11$&$8.49\pm0.08$&
$1400-1450$&$180-250$& $1670\pm30$&$315\pm35$& $1960\pm25$&$195\pm60$&&\\
$f_1$&  $1281.8\pm0.6$&$24.1\pm1.1$& && $1971\pm15$&$240\pm25$& $2310\pm60$&$255\pm70$\\
$h_1$& $1170\pm20$&$360\pm40$& && $1965\pm45$&$345\pm75$& $2215\pm40$&$325\pm55$\\
$b_1$& $1229.5\pm3.2$&$142\pm9$& && $1960\pm35$&$230\pm50$& $2240\pm35$&$320\pm85$\\
$\pi_1$& $1376\pm17$&$300\pm40$& $1653^{+18}_{-15}$&$225^{+45}_{-28}$& $2013\pm25$&$230\pm105$&&\\
$f_2$& $1275\pm1$&$185.1^{+3.5}_{-2.6}$& $1638\pm6$&$99^{+28}_{-24}$& \!\!\!\!\!\!
\begin{tabular}{l}
$1934\pm20$\\
$2001\pm10$\\
\end{tabular}&\!\!\!\!\!\!
\begin{tabular}{l}
$271\pm25$\\
$312\pm32$\\
\end{tabular}&\!\!\!\!\!\!
\begin{tabular}{l}
$2240\pm15$\\
$2293\pm13$\\
\end{tabular}&\!\!\!\!\!\!
\begin{tabular}{l}
$241\pm30$\\
$216\pm37$\\
\end{tabular}\\
$\pi_2$&&& $1672\pm3$&$259\pm9$& $2005\pm15$&$200\pm40$& $2245\pm60$&$320^{+100}_{-40}$\\
$\eta_2$&&& $1617\pm5$&$181\pm11$& $2030\pm16$&$205\pm18$& $2267\pm14$&$290\pm50$\\
$a_2$& $1318.3\pm0.6$&$107\pm5$& $1732\pm16$&$194\pm40$& $2030\pm20$&$205\pm30$& %$2175\pm40$&&
$2255\pm20$&$230\pm15$\\
$\rho_2$&&&&& $1940\pm40$&$155\pm40$& $2225\pm35$&$335^{+100}_{-50}$\\
$\omega_2$&&&&& $1975\pm20$&$175\pm25$& $2195\pm30$&$225\pm40$\\
$f_3$&&&&& $2048\pm8$&$213\pm34$& $2303\pm15$&$214\pm29$\\
$\omega_3$&&& $1667\pm4$&$168\pm10$& $1945\pm20$&$115\pm22$& \!\!\!\!\!\!
\begin{tabular}{l}
$2255\pm15$\\
$2285\pm60$\\
\end{tabular}&\!\!\!\!\!\!
\begin{tabular}{l}
$175\pm30$\\
$230\pm40$\\
\end{tabular}\\
$\rho_3$&&& $1688\pm2.1$&$161\pm10$& $1982\pm14$&$188\pm24$& $2260\pm20$&$160\pm25$\\
$a_3$&&&&& $2031\pm12$&$150\pm18$& $2275\pm35$&$350^{+100}_{-50}$\\
$h_3$&&&&& $2025\pm20$&$145\pm30$& $2275\pm25$&$190\pm45$\\
$b_3$&&&&& $2032\pm12$&$117\pm11$& $2245\pm50$&$320\pm70$\\
$\pi_4$&&&&&&& $2250\pm15$&$215\pm25$\\
$f_4$&&&&& $2018\pm6$&$182\pm7$& $2283\pm17$&$310\pm25$\\
$\rho_4$&&&&&&& $2230\pm25$&$210\pm30$\\
$a_4$&&&&& $2005^{+25}_{-45}$&$180\pm30$& $2255\pm40$&$330^{+110}_{-50}$\\
$\eta_4$&&&&&&& $2328\pm38$&$240\pm90$\\
$\omega_5$&&&&&&& $2250\pm70$&$320\pm95$\\
$\rho_5$&&&&&&& $2300\pm45$&$260\pm75$\\
\noalign{\smallskip}\hline
\noalign{\smallskip}
$\bar M$&$1325\!\pm89\!\pm\!31$&&$1697\!\pm\!56\!\pm\!12$&&$2004\!\pm\!40\!\pm\!24$&&$2269\!\pm\!37\!\pm\!32$&\\
\noalign{\smallskip}
$\bar \Gamma$&&$248\!\pm\!132\!\pm\!57$&&$199\!\pm\!66\!\pm\!29$&&$224\!\pm\!69\!\pm\!38$&&$266\!\pm\!56\!\pm\!53$\\
\noalign{\smallskip}
$\bar \Gamma/\bar M$&&$0.187$&&$0.117$&&$0.112$&&$0.117$\\
%\noalign{\smallskip}
\noalign{\smallskip}\hline
\end{tabular}
% Or use
%\vspace*{5cm}  % with the correct table height
\end{table*}

%\begin{table}
%\caption{Comparison of the exponential and polynomial interpolations for deviations from equidistant clusters %(in MeV).}
%\label{tab:2}
%\begin{tabular}{lllll}
%\hline\noalign{\smallskip}
%$n$ & 1 & 2 & 3 & 4\\
%\noalign{\smallskip}\hline\noalign{\smallskip}
%$\delta_{\text{e}}(n)$ & 89 & 56 & 37 & 26\\
%$\delta_{\text{p}}(n)$ & 89 & 56 & 40 & 31\\
%$\delta_{\text{exper}}(n)$ & 89 & 56 & 40 & 37\\
%\noalign{\smallskip}\hline
%\end{tabular}
%\end{table}


\begin{thebibliography}{99}
\bibitem{rem} A curious historical remark: Balmer guessed the spectrum of excitations 
for the hydrogen atom just 
scrutinizing the available experimental data. Subsequently, the derivation of this spectrum within
the Bohr model marked a milestone in the foundation of Quantum Mechanics. 
At present we may be close to Balmer's situation in the case of light meson spectrum...
\end{thebibliography}
\end{document}